\documentclass[10pt,english,floatfix,nofootinbib,superscriptaddress,aps,prd,preprint, twocolumn]{revtex4}
\usepackage[utf8]{inputenc}
\usepackage{float}
\usepackage{array}
\usepackage{lipsum}
\usepackage{bbm}
\usepackage{dsfont}
\usepackage{graphicx}
\usepackage{amsmath}
\usepackage{tikz}
\usepackage{multirow}
\usepackage{braket}
\usepackage{bbm}
 \usetikzlibrary{quotes,angles}
 \usetikzlibrary{arrows} 
\usetikzlibrary{decorations.markings}
\usepackage{graphicx}
\usepackage[english]{babel}
\usepackage{color}
\usepackage{subfig}
\usepackage{caption}
\usepackage{tensor}
\usepackage{esint}
\usepackage[dvips]{epsfig}
\usepackage[dvips]{graphicx}
\usepackage{float}
\usepackage{units}
\usepackage{textcomp}
\usepackage{mathrsfs}
\usepackage{amsmath}
\usepackage[makeroom]{cancel}
\usepackage{amssymb}
\usepackage{amsbsy}
\usepackage{amsfonts}
\usepackage{amssymb,mathrsfs,xcolor}
\usepackage{esint}
\usepackage{array}
\usepackage{graphicx}

\usepackage{hyperref}
\hypersetup{
    colorlinks,
    citecolor=blue,
    filecolor=green,
    linkcolor=purple,
    urlcolor=red,
}

\usepackage{slashed}

\newcommand{\ie}{\begin{equation}}
\newcommand{\fe}{\end{equation}}
\newcommand{\se}{\begin{eqnarray}}
\newcommand{\ff}{\end{eqnarray}}

\usepackage{hyperref}
\hypersetup{colorlinks,breaklinks,
			citecolor=[rgb]{0,0.0,1.0},
            urlcolor=[rgb]{0.0,0.0,1.0},
            linkcolor=[rgb]{0,0.5,0.9}}

\begin{document}

\title{Influence of a Kalb--Ramond black hole on neutrino behavior}

\author{Yuxuan Shi}
\email{shiyx2280771974@gmail.com}
\affiliation{Department of Physics, East China University of Science and Technology, Shanghai 200237, China}

\author{A. A. Ara\'{u}jo Filho}
\email{dilto@fisica.ufc.br}
\affiliation{Departamento de Física, Universidade Federal da Paraíba, Caixa Postal 5008, 58051--970, João Pessoa, Paraíba,  Brazil.}
\affiliation{Departamento de Física, Universidade Federal de Campina Grande Caixa Postal 10071, 58429-900 Campina Grande, Paraíba, Brazil}


\date{\today}

\begin{abstract}

In this work, we investigate the consequences of Lorentz symmetry breaking induced by a black hole solution within Kalb--Ramond gravity on neutrino dynamics. The analysis is structured around three key phenomena: the energy output associated with neutrino--antineutrino annihilation, the impact of spacetime geometry on the oscillation phases of neutrinos, and the changes in flavor transition probabilities caused by gravitational lensing effects. To complement the theoretical framework, we also carry out a numerical evaluation of neutrino oscillation probabilities, comparing scenarios of normal and inverted mass orderings within a two--flavor approximation.

\end{abstract}

\maketitle


\section{Introduction}

The principle that physical laws keep unchanged under transformations between inertial frames—known as Lorentz symmetry—has been rigorously confirmed through a wide range of experimental tests. However, some theoretical frameworks predict potential violations of this symmetry in high--energy or strong--field regimes. Various studies have explored such possibilities across multiple frameworks \cite{1,5,4,2,3,6,8,7,ghosh2023does,heidari2023gravitational,araujo2023thermodynamics}. The so--called Lorentz symmetry breaking is typically categorized as either explicit or spontaneous \cite{bluhm2006overview}. In its explicit version, symmetry--violating terms are introduced directly into the dynamical equations, resulting in observable anisotropies. On the other hand, its spontaneous approach occurs when the underlying equations preserve Lorentz invariance, but the vacuum state does not, leading to symmetry violation emerging from the structure of the ground state itself \cite{bluhm2008spontaneous}.

Alternative gravitational models that incorporate spontaneous Lorentz symmetry breaking have drawn increasing attention, particularly within the broader scope of the Standard Model Extension \cite{13,heidari2024scattering,AraujoFilho:2024ykw,KhodadiPoDU2023,11,liu2024shadow,filho2023vacuum,9,10,12}. Among these, bumblebee frameworks stand out by introducing a vector field that spontaneously acquires a non--zero vacuum expectation value, which results in preferred spacetime directions and a consequent violation of local Lorentz invariance. Rather than being imposed at the level of fundamental equations, this symmetry breaking emerges from the vacuum structure alone \cite{bluhm2006overview}. These models have been employed to investigate a variety of gravitational phenomena, with particular emphasis on how the symmetry--breaking mechanism affects the thermodynamic behavior of black holes and related systems \cite{reis2021thermal,paperrainbow,araujo2021thermodynamic,aa2021lorentz,araujo2021higher,araujo2022does,anacleto2018lorentz,aaa2021thermodynamics,petrov2021bouncing2,araujo2022thermal,aa2022particles}.

Static and spherically symmetric configurations within the context of bumblebee gravity were initially proposed in Ref. \cite{14}. Since then, modified Schwarzschild--type metrics derived from this framework have been studied extensively across a range of physical applications. These investigations take into account the dynamics of matter accreting onto compact objects \cite{18,17}, the analysis of gravitational lensing phenomena \cite{15}, mechanisms associated with Hawking radiation \cite{kanzi2019gup} and the spectrum of quasinormal oscillations \cite{19,Liu:2022dcn}. Additionally, recent studies have addressed the influence of such backgrounds on quantum processes, including particle production near black holes \cite{araujo2025does}.

Studies aiming to generalize black hole configurations beyond conventional models have explored several modifications to the (A)dS--Schwarzschild geometry \cite{Bailey:2025oun}. One such modification involves relaxing the usual vacuum conditions, which results in altered spacetime dynamics \cite{20}. This idea was later extended by considering black hole solutions within bumblebee gravity, where a background vector field develops a non--zero component along the time direction, thereby inducing spontaneous Lorentz symmetry breaking \cite{23,22,24,21}.

A distinct approach to implementing Lorentz symmetry violation makes use of the Kalb--Ramond framework, where an antisymmetric tensor field of rank two, originally derived from bosonic string theory, plays a fundamental role \cite{43,42,maluf2019antisymmetric}. When this field is non--minimally coupled to gravity and develops a non--zero vacuum expectation value, it triggers spontaneous breaking of Lorentz invariance under particle transformation. A static and spherically symmetric black hole solution under these conditions was explored in \cite{44}, with further studies addressing the influence of this background on test particle trajectories \cite{45}. Investigations have also extended to the rotating case, particularly focusing on how such configurations affect light deflection and shadow structures \cite{46}.

In contrast to most elementary particles, neutrinos exhibit a peculiar behavior that has made them a central topic in modern particle physics research \cite{neu44,neu43,neu42}. Their interaction states—identified by flavor—do not align with their physical mass states, a property that leads to a nontrivial quantum evolution. As neutrinos travel through space, this feature results in a periodic transition between flavors, a phenomenon termed neutrino oscillation, rooted in the interference between distinct mass eigenstates \cite{neu40,neu39,neu41}.

Neutrino oscillation patterns, when described in the context of flat spacetime, are dictated by the differences between the squared masses of the eigenstates rather than the absolute mass values. These differences are defined as  
\[
\Delta m^2_{ij} = m_i^2 - m_j^2,
\]  
with commonly used parameters including $|\Delta m^2_{21}|$, $|\Delta m^2_{31}|$, and $|\Delta m^2_{23}|$. The oscillation probabilities derived from this framework depend solely on these mass--squared gaps, making it impossible to extract the exact masses of the neutrinos directly from oscillation data \cite{neu45}.

The behavior of neutrino oscillations is significantly altered in the presence of gravitational fields. Unlike the flat spacetime case—where only mass--squared differences influence flavor transitions—spacetime curvature can affect the phase evolution in a way that introduces sensitivity to the absolute neutrino masses \cite{AraujoFilho:2025rzh,Shi:2024flw}. In curved geometries, additional contributions emerge in the oscillation phase, modifying the standard formulas. These gravitational corrections are particularly relevant for ultra--relativistic neutrinos originating from distant cosmic sources \cite{Shi:2023kid}. The flavor distribution of neutrinos detected after long--distance travel through curved spacetime can serve as a probe of both their intrinsic properties and the gravitational fields encountered along their trajectories. By contrasting observational data with predictions obtained from neutrino oscillation models adapted to curved backgrounds, one can identify deviations that reflect the influence of spacetime curvature. This comparison provides a means to investigate not only the neutrino mass structure but also the characteristics of the gravitational environments through which they have passed \cite{neu50,neu53,Shi:2023hbw,neu48,neu52,neu47,neu46,neu49,neu51}. 

Neutrinos are superior to electromagnetic probes in several respects, particularly in environments with strong gravitational fields. Unlike photons, they can traverse dense regions—such as the inner cores of supernovae, accretion disks, or the vicinity of compact objects—where electromagnetic signals are typically absorbed or scattered. Their weak interactions reduce decoherence effects, allowing them to carry largely undistorted information from otherwise inaccessible regions. Moreover, the phase evolution of neutrino flavor states encodes both mass differences and features of the surrounding spacetime geometry through quantum interference, making them sensitive probes of curvature. In extreme scenarios, such as gamma-ray bursts, ultra--high--energy neutrinos may even penetrate regions that photons cannot reach, offering unique access to the deep structure of spacetime.

The structure of spacetime is fundamental in shaping neutrino flavor evolution during propagation. When analyzed geometrically, the phase accumulation along neutrino geodesics becomes sensitive to the underlying gravitational field \cite{neu55,neu54}. In strongly curved regions—especially around dense astrophysical sources—gravitational lensing can bend neutrino paths, leading to convergence or overlap of trajectories. This deflection influences quantum interference, thereby modifying the oscillation pattern and altering the expected transition probabilities between flavors \cite{Shi:2024flw,neu53}.

Gravitational lensing effects on neutrino oscillations have sparked growing interest, particularly in cases where the bending of trajectories leads to path convergence and modifies interference conditions \cite{neu58,neu57,neu56}. In the presence of rotating gravitational fields, the situation becomes more nuanced. Swami's analysis revealed that the angular momentum of the central mass alters the phase evolution of propagating neutrinos, which can either suppress or enhance oscillation probabilities depending on the spacetime configuration. This influence becomes especially relevant in astrophysical environments with scales comparable to that of the Sun \cite{neu59}.

Furthermore, neutrino oscillations have also been studied in geometries that deviate from perfect spherical symmetry. In particular, axially symmetric spacetimes parameterized by a deformation factor $\gamma$ have been shown to affect flavor transitions, even when the background remains static and asymptotically flat. The introduction of such geometric deformation alters the neutrino phase evolution in a way that can generate sensitivity to the absolute mass scale—an effect not encountered in standard flat spacetime treatments \cite{neu60}.

In this manner, we explore how neutrino dynamics are affected by Lorentz symmetry violation resulting from a black hole solution formulated within Kalb--Ramond gravity. The discussion begins with the energy production linked to neutrino--antineutrino annihilation near the gravitational source. It then turns to how the modified spacetime geometry influences the phase evolution of neutrino states, followed by an investigation of how gravitational lensing alters flavor conversion rates. To reinforce the analytical treatment, we numerically compute oscillation probabilities in both normal and inverted mass hierarchies, employing a two--flavor approximation.

\section{The black hole in Kalb--Ramond gravity}

Before presenting the black hole solution itself, we briefly remark that Lorentz violation can also be studied directly within the neutrino sector, without resorting to a modified gravitational background. This alternative approach has been extensively explored within the framework of the Standard--Model Extension, where Lorentz--violating effects emerge from modified dispersion relations for neutrinos. Seminal contributions in this context include the works of Kostelecký and Mewes \cite{Kostelecky:2003cr,Kostelecky:2011gq}, and Katori et al. \cite{Katori:2006mz}. While our investigation focuses on spontaneous Lorentz violation induced by a gravitational background, both approaches explore distinct mechanisms through which Lorentz symmetry may be broken and should be viewed in a complementary manner.

A new black hole solution was recently introduced in the literature \cite{yang2023static}, and is given by the following metric:
\ie
\begin{split}
\label{mainmetric}
\mathrm{d}s^{2}  = & - \left( \frac{1}{1-\ell} - \frac{2M}{r}   \right) \mathrm{d}t^{2} + \frac{\mathrm{d}r^{2}}{\frac{1}{1-\ell} - \frac{2M}{r} } \\
& + r^{2}\mathrm{d}\theta^{2} + r^{2} \sin^{2}\mathrm{d}\varphi^{2},
\end{split}
\fe
where $\ell = \xi_{2} \tilde{b}/2$, with $\tilde{b}$ denoting the norm of the vacuum expectation value, defined as $b^{\mu\nu} b_{\mu\nu} \equiv b^{2}$ \cite{yang2023static}.

Several investigations have recently focused on this black hole model, addressing a variety of physical aspects. Research has analyzed the greybody factors \cite{guo2024quasinormal}, lensing behavior \cite{junior2024gravitational}, and quasinormal spectra \cite{araujo2024exploring}. In addition, constraints related to spontaneous symmetry breaking have also been discussed \cite{junior2024spontaneous}, along with studies on orbital dynamics and quasi--periodic oscillations near the event horizon \cite{jumaniyozov2024circular}. The particle production effects were treated in \cite{araujo2024particle}, while the accretion of Vlasov--type matter has been examined in \cite{jiang2024accretion}. Moreover, a generalization involving electric charge has been constructed \cite{duan2024electrically}, and its consequences explored in several follow--up works \cite{hosseinifar2024shadows,al-Badawi:2024pdx,heidari2024impact,chen2024thermal,Zahid:2024ohn,aa2024antisymmetric}. In parallel, an alternative solution inspired by Kalb–Ramond gravity was proposed in \cite{Liu:2024oas}; however, a detailed investigation of this geometry is beyond the objectives of the present analysis and will not be addressed here.

The following section is devoted to examining how the Kalb--Ramond black hole proposed in Ref. \cite{yang2023static} affects various aspects of neutrino dynamics. We begin by analyzing the energy deposition rate resulting from neutrino--antineutrino annihilation. Subsequently, we explore the modifications introduced by the spacetime geometry to the oscillation phase and the corresponding transition probabilities. Once these foundational elements are established, we turn to the study of gravitational lensing effects on neutrino trajectories. To close the section, we present numerical results that complement and support the theoretical framework developed throughout the analysis.


\section{Annihilation--induced energy deposition rate.}

This part of the analysis centers on the mechanism of energy transfer in a spacetime background altered by the Lorentz violating parameter $\ell$, as shown in Eq. (\ref{mainmetric}). Here, the dominant contribution to energy release arises from the annihilation of neutrino--antineutrino pairs. The corresponding energy deposition rate, defined per unit volume and per unit time, can be expressed as \cite{Salmonson:1999es}:
\ie
\dfrac{\mathrm{d}E(r)}{\mathrm{d}t\mathrm{d}V}=2KG_{f}^{2}f(r)\iint
n(\varepsilon_{\nu})n(\varepsilon_{\overline{\nu}})
(\varepsilon_{\nu} + \varepsilon_{\overline{\nu}})
\varepsilon_{\nu}^{3}\varepsilon_{\overline{\nu}}^{3}
\mathrm{d}\varepsilon_{\nu}\mathrm{d} \varepsilon_{\overline{\nu}}
\fe
in which
\ie
K = \dfrac{1}{6\pi}(1\pm4\sin^{2}\theta_{W}+8\sin^{4} \theta_{W}).
\fe
Using the value $\sin^{2}\theta_{W} = 0.23$ for the Weinberg angle, one can determine the specific forms of the energy deposition rates associated with various neutrino–antineutrino pair channels. These expressions, originally derived in \cite{Salmonson:1999es}, reflect the dependence of the process on the weak interaction parameters and the flavor configuration of the annihilating neutrinos
\ie
K(\nu_{\mu},\overline{\nu}_{\mu}) = K(\nu_{\tau},\overline{\nu}_{\tau})
=\dfrac{1}{6\pi}\left(1-4\sin^{2}\theta_{W} + 8\sin^{4}\theta_{W}\right)
\fe
and
\ie
K(\nu_{e},\overline{\nu}_{e})
=\dfrac{1}{6\pi}\left(1+4\sin^{2}\theta_{W} + 8\sin^{4}\theta_{W}\right).
\fe

Each neutrino pair channel yields a specific expression, based on the adopted value of the Weinberg angle, $\sin^{2}\theta_{W} = 0.23$. The interaction strength involved in these weak processes is set by the Fermi constant, given as $G_{f} = 5.29 \times 10^{-44} \, \text{cm}^2 \, \text{MeV}^{-2}$. With these parameters in place, the contribution resulting from the angular integration is expressed as follows \cite{Salmonson:1999es}:
\begin{align}
f(r)&=\iint\left(1-\bm{\Omega_{\nu}}\cdot\bm{\Omega_{\overline{\nu}}}\right)^{2}
\mathrm{d}\Omega_{\nu}\mathrm{d}\Omega_{\overline{\nu}}\notag\\
&=\dfrac{2\pi^{2}}{3}(1 - x)^{4}\left(x^{2} + 4x + 5\right)
\end{align}
where
\ie
x = \sin\theta_{r}.
\fe

The angle $\theta_r$, defined at a radial coordinate $r$, characterizes the deviation between a particle’s motion and the tangential direction of a circular orbit at that location. The propagation of neutrinos and antineutrinos is described by the unit vectors $\Omega_{\nu}$ and $\Omega_{\overline{\nu}}$, while their corresponding directions are integrated over the solid angles $\mathrm{d}\Omega_{\nu}$ and $\mathrm{d}\Omega_{\overline{\nu}}$. It is worthy to be mentioned that, when the system is in thermal equilibrium at a given temperature $T$, the distribution of neutrinos and antineutrinos in phase space, represented by $n(\varepsilon_{\nu})$ and $n(\varepsilon_{\overline{\nu}})$, is governed by the Fermi--Dirac statistics \cite{Salmonson:1999es}
\ie
n(\varepsilon_{\nu}) = \frac{2}{h^{3}}\dfrac{1}{e^{\left({\frac{\varepsilon_{\nu}}{k \, T}}\right)} + 1}.
\fe

Within this framework, Planck’s constant $h$ and Boltzmann’s constant $k$ enter as fundamental parameters. Once these constants are set, one arrives at the expression that determines the energy deposition rate—defined per unit volume and per unit time—associated with the neutrino annihilation process \cite{Salmonson:1999es}
\ie
\frac{\mathrm{d}E}{\mathrm{d}t\mathrm{d}V} = \frac{21\zeta(5)\pi^{4}}{h^{6}}K G_{f}^{2} f(r)(k \, T)^{9}.
\fe
The quantity $\mathrm{d}E/\mathrm{d}t , \mathrm{d}V$ is essential for describing energy conversion processes in the vicinity of compact astrophysical sources \cite{Salmonson:1999es}. This expression reflects how various physical properties vary with position, with particular emphasis on the temperature function $T(r)$, which determines the thermal conditions at each radial coordinate \cite{Salmonson:1999es}.

The temperature measured locally by an observer positioned at a radial coordinate $r$ is affected by the gravitational redshift and satisfies the relation $T(r)\sqrt{-g_{tt}(r)} = \text{constant}$, illustrating how spacetime curvature influences thermal observations \cite{Salmonson:1999es}. At the boundary of the neutrinosphere—defined at $r = R$—the temperature associated with neutrino emission satisfies the condition \cite{Salmonson:1999es}:  
\ie
T(r)\sqrt{-g_{tt}(r)} = T(R)\sqrt{-g_{tt}(R)}.
\fe  
In this expression, $R$ represents the radius of the compact object generating the gravitational field. For ease of calculation, the temperature profile $T(r)$ is re-expressed using the above identity. Incorporating the effects of redshift, the resulting expression for neutrino luminosity takes the following form \cite{Salmonson:1999es}:
\ie
L_{\infty} = -g_{tt}(R)L(R).
\fe

Furthermore, for an individual neutrino flavor, the luminosity calculated at the neutrinosphere is expressed through the following relation \cite{Salmonson:1999es}:
\ie
L(R) = 4 \pi R_{0}^{2}\dfrac{7}{4}\dfrac{a\,c}{4}T^{4}(R).
\fe

The parameter $c$ stands for the speed of light in vacuum and, while $a$ represents the radiation constant. The temperature profile, as observed from a specific radial coordinate, can be reformulated by incorporating gravitational redshift corrections. This leads to the following relation, which correlates the local temperature to the geometry of spacetime \cite{Salmonson:1999es}:
\ie
\begin{split}
\frac{\mathrm{d}E(r)}{\mathrm{d}t \, \mathrm{d}V} & = \dfrac{21\zeta(5)\pi^{4}}{h^{6}}
KG_{f}^{2}k^{9}\left(\dfrac{7}{4}\pi a\,c\right)^{-\frac{9}{4}}\\
& \times L_{\infty}^{\frac{9}{4}}f(r)
\left[\dfrac{\sqrt{-g_{tt}(R)}}{-g_{tt}(r)}\right]^{\frac{9}{2}} R^{-\frac{9}{2}}.
\end{split}
\fe

Notice that the function $\zeta(s)$ appearing in the above expression corresponds to the Riemann zeta function, which for real values of $s$ greater than one is given by the infinite series representation:
\ie
\zeta(s) = \sum_{n=1}^{\infty} \frac{1}{n^s},
\fe

It should be emphasized that the local energy deposition rate is not solely governed by the radial dependence; it also incorporates the influence of the metric components evaluated at the surface of the gravitational source. To compute the total radiative energy output under the effects of spacetime curvature, one must perform a time integration of the local deposition density. The evaluation of the angular factor $f(r)$ requires a more detailed analysis of the variable $x$, previously introduced in the context of angular integration. To do so, we write \cite{Salmonson:1999es,Lambiase:2020iul,Shi:2023kid,AraujoFilho:2024mvz}
\begin{align}
x^{2}& = \sin^{2}\theta_{r}|_{\theta_{R}=0}\notag\\
&=1-\dfrac{R^{2}}{r^{2}}\dfrac{g_{tt}(r)}{g_{tt}(R)}.
\end{align}

The total energy transferred to the surrounding region can be obtained by integrating the local deposition rate—given per unit volume and per unit time—over the spatial domain influenced by the gravitational source. This integration relies on the angular factor, which is intrinsically linked to the properties of the underlying spacetime geometry. The structure of the metric directly governs how this angular term behaves \cite{Lambiase:2020iul,Shi:2023kid}
\ie
\begin{split}
\dot{Q} & = \frac{\mathrm{d}E}{\sqrt{-g_{tt}(r)}\mathrm{d}t}\\
&=\dfrac{84\zeta(5)\pi^{5}}{h^{6}}KG_{f}^{2}k^{9}
\left(\dfrac{7}{4}\pi a\,c\right)^{-\frac{9}{4}}
L_{\infty}^{\frac{9}{4}}\left[-g_{tt}(R)\right]^{\frac{9}{4}}\\
& \times 
R^{-\frac{3}{2}}\int_{1}^{\infty}(x-1)^4\left(x^2+4x+5\right)\dfrac{y^2}{\left[-g_{tt}(yR)\right]^5}\mathrm{d}y.
\end{split}
\fe

The quantity $\dot{Q}$ represents the overall energy conversion rate from neutrinos into electron–positron pairs at a given radial location \cite{Salmonson:1999es}. When this rate reaches high enough values, it may lead to intense pair production capable of initiating astrophysical events. For the sake of obtaining a better understanding in such a direction, it is essential to contrast this relativistic expression for energy deposition with the corresponding Newtonian formulation, highlighting the role of gravitational corrections \cite{Salmonson:1999es,Lambiase:2020iul,Shi:2023kid,shi2022neutrino}
\ie
\begin{split}
\label{ratio_Q}
 \frac{\dot{Q}}{\dot{Q}_{\text{Newton}}} = 3\left[-g_{tt}(R)\right]^{\frac{9}{4}}
& \int_{1}^{\infty}(x - 1)^{4}\left(x^{2} + 4x + 5\right) \\
& \times \frac{y^{2}}{\left[-g_{tt}(yR)\right]^5}\mathrm{d}y.
\end{split}
\fe
with
\ie
\begin{split}
g_{tt}(R)&= -\left(\dfrac{1}{1-\ell} - \frac{2M}{R}\right),\\
g_{tt}(yR)&= -\left(\dfrac{1}{1-\ell} - \frac{2M}{yR}\right).
\end{split}
\fe
Additionally, we also have
\begin{align}
x^{2}=1-\dfrac{1}{y^{2}}\frac{\dfrac{1}{1-\ell}-\dfrac{2M}{yR}}{\dfrac{1}{1-\ell}-\dfrac{2M}{R}}.
\end{align}

\begin{figure}
    \centering
    \includegraphics[scale=0.55]{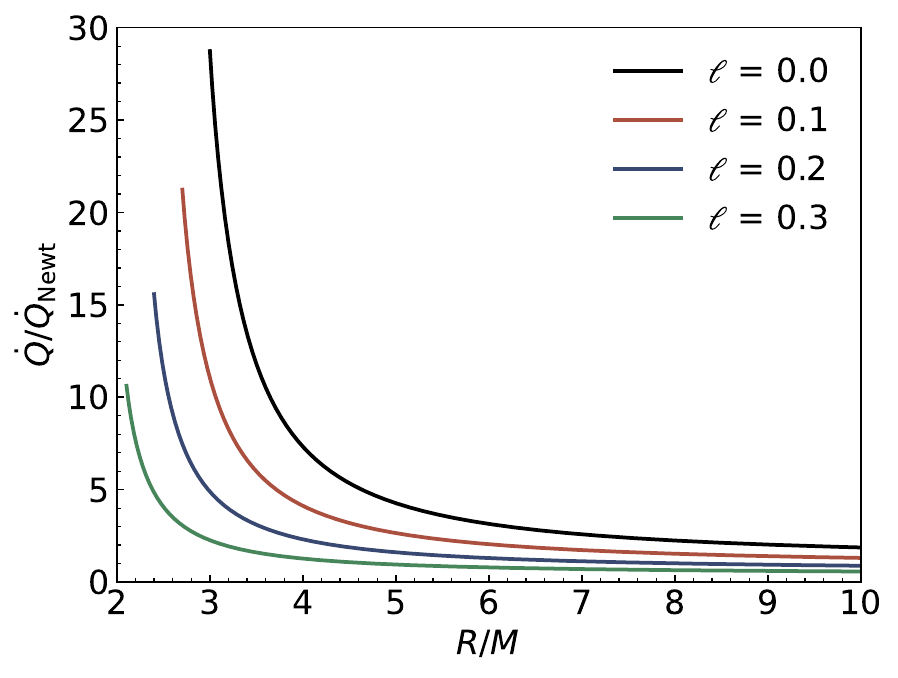}
    \caption{The ratio $\dot{Q}/\dot{Q}_{\text{Newton}}$ is plotted against $R/M$ for various choices of the parameter $\ell$, illustrating how deviations from the Newtonian case depended on both the radial distance and the Lorentz--violating contribution.}
    \label{esndedrdgyddepodsiidtodn}
\end{figure}

To evaluate the impact of Kalb--Ramond gravitational corrections on neutrino--induced energy deposition, we analyze the expression in Eq.~(\ref{ratio_Q}) and present the corresponding ratio profiles in Fig.~\ref{esndedrdgyddepodsiidtodn}, considering several values of the parameter $\ell$. Despite changes in $\ell$, the qualitative behavior of the curves remains similar. The results demonstrate that the presence of Kalb--Ramond contributions suppresses the conversion efficiency of neutrino–antineutrino pairs into electron–positron pairs. For instance, at $R = 10M$, the ratios fall below a factor of two compared to the classical framework, and at $R = 3M$, they are approximately reduced by half.

Assuming a neutrino luminosity at spatial infinity of roughly $\sim10^{53}\,\mathrm{erg/s}$, with the source radius fixed at $R = 20\,\mathrm{km}$ and the parameter $D = 1.23$, Tab~\ref{tab:dotQ} provides a comparative account of the energy deposition rate. The results clearly show that the inclusion of Kalb--Ramond corrections leads to a notable reduction in the total energy output compared to the standard case.

\begin{figure}
    \centering
    \includegraphics[scale=0.55]{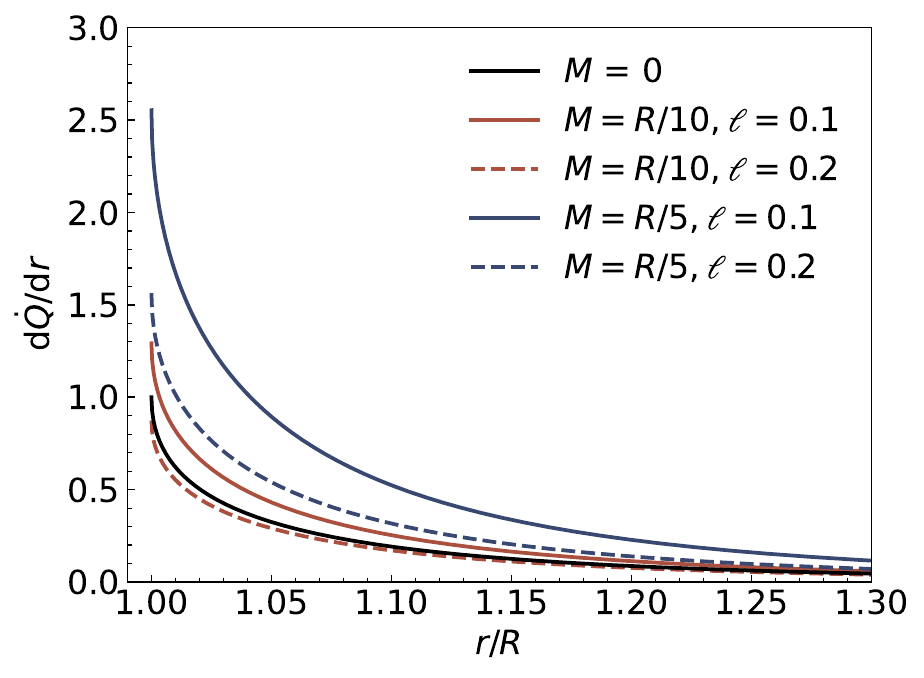}
    \caption{The behavior of $\mathrm{d}\dot{Q}/\mathrm{d}r$ is examined for several values of the ratio $M/R$. In the Newtonian limit, where $M = 0$, the expression simplifies, yielding $\mathrm{d}\dot{Q}/\mathrm{d}r = 1$ at the surface $r = R$.}
    \label{dQdr}
\end{figure}

This reduction is further characterized by the radial behavior of the deposition rate derivative, $\mathrm{d}\dot{Q}/\mathrm{d}r$, shown in Fig.~\ref{dQdr} for various compactness ratios $M/R$. It is important highlight that this quantity describes the radial dependence of energy deposition and encodes the geometric effects of spacetime via the metric functions. Examining how modifications to gravity—specifically those introduced by the Kalb--Ramond field—affect neutrino-–antineutrino annihilation is crucial for determining the circumstances under which such mechanisms could drive phenomena such as gamma--ray bursts. Additionally, it is worth noting that a similar analysis was performed for a rotating black hole within the framework of Kalb--Ramond gravity as well \cite{Khodadi:2023yiw}.

\begin{table}[h!]
\centering
\caption{The energy emission rate $\dot{Q}$, measured in erg/s, is evaluated across various values of the Lorentz--violating parameter $\ell$ and the dimensionless ratio $R/M$.}
\label{tab:dotQ}
\begin{tabular}{ccc}
\hline\hline
$\ell$ & $R/M$ & $\dot{Q}$ \((\text{erg/s})\) \\
\hline
0.0 & 0 & $1.50 \times 10^{50}$ \\
\hline
\multirow{2}{*}{0.0} & 3 & $4.32 \times 10^{51}$ \\
                      & 4 & $1.10 \times 10^{51}$ \\
\hline
\multirow{2}{*}{0.1} & 3 & $1.66 \times 10^{51}$ \\
                      & 4 & $0.62 \times 10^{51}$ \\
\hline
\multirow{2}{*}{0.2} & 3 & $0.73 \times 10^{51}$ \\
                      & 4 & $0.35 \times 10^{51}$ \\
\hline
\multirow{2}{*}{0.3} & 3 & $0.34 \times 10^{51}$ \\
                      & 4 & $0.19 \times 10^{51}$ \\
\hline\hline
\end{tabular}
\end{table}


\section{Neutrino phase evolution and conversion probability}

The motion of neutrinos associated with the $k$--th eigenmode, within a spherically symmetric background described by the metric in Eq. (\ref{mainmetric}), can be analyzed by employing the Lagrangian formalism developed in Ref.~\cite{neu18}
\begin{align}
\mathcal{L}
& = -\frac{1}{2}  m_{k} g_{tt}(r)\left(\frac{\mathrm{d}t}{\mathrm{d}\tau}\right)^2-\frac{1}{2}m_{k}g_{rr}(r)\left(\frac{\mathrm{d}r}{\mathrm{d}\tau}\right)^2 \notag\\
& \quad -\dfrac{1}{2}m_{k}r^2\left(\frac{\mathrm{d}\theta}{\mathrm{d}\tau}\right)^2  
 -\frac{1}{2}m_{k}r^2\sin^2\theta\left(\frac{\mathrm{d}\varphi}{\mathrm{d}\tau}\right)^2.
\end{align}

When the motion is restricted to the equatorial plane, specified by $\theta = \pi/2$, only a subset of the momentum components remain nonzero. These components can be derived by first defining the canonical momentum $p_{\mu}$ as the partial derivative of the Lagrangian $\mathcal{L}$ with respect to the generalized velocity, i.e., $p_{\mu} = \partial \mathcal{L} / \partial (\mathrm{d}x^{\mu} / \mathrm{d}\tau)$. Here, $\tau$ stands for the proper time, while $m_k$ corresponds to the mass of the neutrino in the $k$--th mass eigenstate. The explicit expressions for these non--vanishing components can be found in Refs.~\cite{neu60,Shi:2024flw}
\begin{align}
p^{(k)t} &= -m_{k}g_{tt}(r)\frac{\mathrm{d}t}{\mathrm{d}\tau} = -E_{k}, \\
p^{(k)r} &= m_{k}g_{rr}(r)\frac{\mathrm{d}r}{\mathrm{d}\tau}, \\
p^{(k)\varphi} &= m_{k}r^2\frac{\mathrm{d}\varphi}{\mathrm{d}\tau} = J_{k},
\end{align}
with the mass of the $k$--th eigenstate obeying the mass--shell condition, which imposes the constraint $g^{\mu\nu} p_\mu p_\nu = -m_k^2$, ensuring naturally consistency with the particle's relativistic dynamics in curved spacetime \cite{neu54,neu55}
\begin{align}
-m_{k}^2 =g^{tt}p_t^2+g^{rr}p_r^2+g^{\varphi\varphi}p_{\varphi}^2.
\end{align}

In regions where gravitational effects are small, analyses of neutrino oscillations have commonly adopted the plane wave approximation \cite{neu54,neu53}. As weak interactions govern both the emission and detection processes, neutrinos are observed in flavor eigenstates rather than in states of definite mass, as discussed in \cite{neu62,neu61,neu63,Shi:2024flw}
\begin{align}
\ket{\nu_{\alpha}} = \sum \mathrm{U}_{\alpha i}^{*}\ket{\nu_{i}}.
\end{align}

Neutrino propagation is best described in terms of mass eigenstates, denoted by $\ket{\nu_i}$, each of which evolves distinctly across spacetime. However, neutrinos are created and detected in flavor states—electron, muon, and tau—indexed by $\alpha = e, \mu, \tau$. The transformation between these two representations involves a $3 \times 3$ unitary matrix $\mathrm{U}$, as established in \cite{neu41}. To track this evolution, one assigns the spacetime coordinates $\left(t_{S}, \bm{x}_{S}\right)$ and $\left(t_{D}, \bm{x}_{D}\right)$ to the source (emission point) and detector (detection point), respectively. The wave function corresponding to each mass eigenstate then propagates along the path connecting these two events
\begin{align}
\ket{\nu_{i}\left(t_{D},\bm{x}_{D}\right)} = \exp({-\mathrm{i}\Phi_{i}})\ket{\nu_{i}\left(t_{S},\bm{x}_{S}\right)},
\end{align}
so that the associated phase acquired by each mass eigenstate along its trajectory is given by the expression
\begin{align}
\Phi_{i}=\int_{\left(t_{S},\bm{x}_{S}\right)}^{\left(t_{D},\bm{x}_{D}\right)}g_{\mu\nu}p^{(i)\mu}\mathrm{d}x^{\nu}.
\end{align}

Furthermore, this analysis approaches flavor oscillation by examining the evolution of a neutrino between its point of origin and the location of detection. As the particle propagates, it may undergo a transformation from one flavor state $\nu_{\alpha}$ to another $\nu_{\beta}$. The likelihood of detecting such a transition is quantified by the probability expression below:
\begin{align}
\mathcal{P}_{\alpha\beta}
& = |\left\langle \nu_{\beta}|\nu_{\alpha}\left(t_{D}, \bm{x}_{D}\right)\right\rangle|^2 \\
& = \sum_{i,j} \mathrm{U}_{\beta i}\mathrm{U}_{\beta j}^{*} \mathrm{U}_{\alpha j} \mathrm{U}_{\alpha i}^{*}\,  \exp{[-\mathrm{i}(\Phi_{i}-\Phi_{j})]}.
\end{align}

The analysis focuses on neutrinos confined to the equatorial plane ($\theta = \pi/2$) within the gravitational field of a Kalb--Ramond black hole. In this scenario, the phase associated with their motion is given by the following expression:
\begin{align}
\label{Pgefhi}
\Phi_{k} & = \int_{\left(t_{S},\bm{x}_{S}\right)}^{\left(t_{D}, \bm{x}_{D}\right)} g_{\mu\nu} p^{(k)\mu}\mathrm{d}x^{\nu}\notag\\
& = \int_{\left(t_{S},\bm{x}_{S}\right)}^{\left(t_{D}, \bm{x}_{D}\right)}\left[E_{k}\mathrm{d}t - p^{(k)r}\mathrm{d}r-J_{k}\mathrm{d}\varphi\right] \notag\\
& = \pm\frac{m_{k}^2}{2E_0}\int_{r_{S}}^{r_{D}}\sqrt{-g_{tt}g_{rr}}\left(1-\dfrac{b^2|g_{tt}|}{g_{\varphi\varphi}}\right)^{-\frac{1}{2}}\mathrm{d}r.
\end{align}

In the case of a weak gravitational field, where $M/r \ll 1$, an expansion of the integrand in Eq.~(\ref{Pgefhi}) is possible
\begin{align}
&\quad\sqrt{-g_{tt}g_{rr}}\left(1-\dfrac{b^2|g_{tt}|}{g_{\varphi\varphi}}\right)^{-\frac{1}{2}}\notag\\
&\simeq\dfrac{1}{\sqrt{1-\frac{b^2}{(1-\ell)r^2}}}+\dfrac{b^2(1-\ell)M}{\sqrt{1-\frac{b^2}{(1-\ell)r^2}}\left[b^2-(1-\ell)r^2\right]r}.
\end{align}
As a result, the phase can be expressed as shown below
\begin{align}
\Phi_k&=\dfrac{m_k^2}{2E_0}\Biggl[\sqrt{r_D^2-\dfrac{b^2}{1-\ell}}-\sqrt{r_S^2-\dfrac{b^2}{1-\ell}}\notag\\
&\quad+(1-\ell)M\left(\dfrac{r_D}{\sqrt{r_D^2-\frac{b^2}{1-\ell}}}-\dfrac{r_S}{\sqrt{r_S^2-\frac{b^2}{1-\ell}}}\right)\Biggr].
\end{align}

In this setup, the mean energy of the relativistic neutrinos emitted from the source is represented by $E_0 = \sqrt{E_k^2 - m_k^2}$, with $E_k$ being the energy of the $k$--th eigenstate and $m_k$ its corresponding mass. The impact parameter, denoted by $b$, is discussed in \cite{neu18}. While traveling through curved spacetime, the neutrinos reach a minimum distance at $r = r_0$. To find this closest approach point in the weak field approximation, the relevant equation can be solved
\begin{align}
\left(\dfrac{\mathrm{d}r}{\mathrm{d}\varphi}\right)_0=\pm\dfrac{g_{\varphi\varphi}}{b^2}\sqrt{\dfrac{1}{-g_{tt}g_{rr}}-\dfrac{b^2}{g_{rr}g_{\varphi\varphi}}}=0.
\end{align}

By solving the orbital equation governing the neutrino’s trajectory in weak gravitational fields, the closest approach distance, $r_0$, is obtained
\begin{align}
\label{r0}
r_0 \simeq \dfrac{b}{\sqrt{1-\ell}}-(1-\ell)M.
\end{align}
The total phase experienced by a neutrino from its emission at the source, through its closest radial approach, to its detection is determined by applying Eq.~(\ref{Pgefhi}) in combination with the formula for $r_0$ provided in Eq.~(\ref{r0}). In this manner, we obtain
\begin{align}
\label{pphhiii}
&\quad\Phi_{k}\left(r_{S}\to r_{0} \to r_{D}\right)\notag\\
&\simeq \frac{{m}_{k}^2}{2E_0}
\Biggl[\sqrt{r_D^2-r_0^2}+\sqrt{r_S^2-r_0^2}\notag\\
&\quad+(1-\ell)M\left(\sqrt{\dfrac{r_D-r_0}{r_D+r_0}+\dfrac{r_S-r_0}{r_S+r_0}}\right)\Biggr],
\end{align}
or, in other words, 
\begin{align}
\Phi_{k}
& \simeq \frac{{m}_{k}^2}{2E_0}
\Biggl\{\sqrt{r_D^2-\dfrac{b^2}{1-\ell}}+\sqrt{r_S^2-\dfrac{b^2}{1-\ell}}\notag\\
&\quad+(1-\ell)M\left[\dfrac{b}{\sqrt{(1-\ell)r_D^2-b^2}}+\dfrac{b}{\sqrt{(1-\ell)r_S^2-b^2}}\right]\notag\\
&\quad+\sqrt{\dfrac{\sqrt{1-\ell}r_D-b}{\sqrt{1-\ell}r_D+b}}+\sqrt{\dfrac{\sqrt{1-\ell}r_S-b}{\sqrt{1-\ell}r_S+b}}\Biggr\}.
\end{align}

Next, we perform an expansion of the preceding formula in powers of $b/r_{S,D}$, retaining terms up to $\mathcal{O}\left(b^2/r_{S,D}^2\right)$, under the condition that $b \ll r_{S,D}$. This yields the following result
\begin{align}
\Phi_k=\dfrac{m_k^2}{2E_0}(r_D+r_S)\left[1-\dfrac{1}{2(1-\ell)}\dfrac{b^2}{r_Dr_S}+\dfrac{2(1-\ell)M}{r_D+r_S}\right].
\end{align}

An increase in the Lorentz--violating parameter $\ell$ leads to a noticeable effect: the phase accumulated by neutrinos during their propagation decreases as $\ell$ grows. For this analysis, the following parameters are chosen: $E_0 = 10\,\mathrm{MeV}$, $r_{D} = 10\,\mathrm{km}$, and $r_{S} = 10^5 r_{D}$.

Within this viewpoint, gravitational lensing plays a role in altering the neutrino trajectories as they pass through curved spacetime. To investigate the flavor oscillation probability near the black hole, it is essential to compute the phase differences along the distinct paths that the neutrinos can take \cite{Shi:2024flw,AraujoFilho:2025rzh}
\begin{align}
\Delta\Phi_{ij}^{pq}
&= \Phi_i^{p}-\Phi_j^{q}\notag\\
&= \left(\Delta m_{ij}^2 A_{pq}+\Delta b_{pq}^2 B_{ij}\right),
\end{align}
in which
\begin{align}
\Delta m_{ij}^2 & = m_i^2 - m_j^2,\\
\Delta b_{pq}^2 & = \frac{b_{p}^2-b_{q}^2}{1-\ell},\\
A_{pq} & = \frac{r_{S} + r_{D}}{2 E_0}\left[1+\dfrac{2(1-\ell)M}{r_D+r_S}-\dfrac{\sum b_{pq}^2}{4r_Dr_S}\right],\\
B_{ij} & = -\frac{\sum m_{ij}^2}{8E_0}\left(\frac{1}{r_{D}} + \frac{1}{r_{S}}\right),\\
\sum b_{pq}^2 & = \dfrac{b_{p}^2 + b_{q}^2}{1-\ell},\\
\sum m_{ij}^2 & = m_i^2 + m_j^2.
\end{align}

Phases corresponding to different neutrino trajectories are labeled using superscripts, such as $\Phi_i^p$, where each $p$ refers to a specific path determined by its associated impact parameter $b_p$. The difference in phase that contributes to the transition probability in a Lorentz--violating black hole background arises from the individual neutrino masses $m_i$, the squared mass differences $\Delta m_{ij}^2$, and the structure of the curved spacetime. As one should naturally expect, when $\ell \to 0$, this phase difference expression reduces to the conventional result presented in Ref.~\cite{neu53}.

It is important to mention that while $B_{ij}$ encapsulates the dependence on neutrino mass parameters, the effect of $\ell$ appears only as a correction to $A_{pq}$, which alters both the magnitude of the acquired phase and the resulting oscillation amplitude.


\section{Gravitational deflection of neutrino trajectories}

\begin{figure}
    \centering
    \includegraphics[scale=0.45]{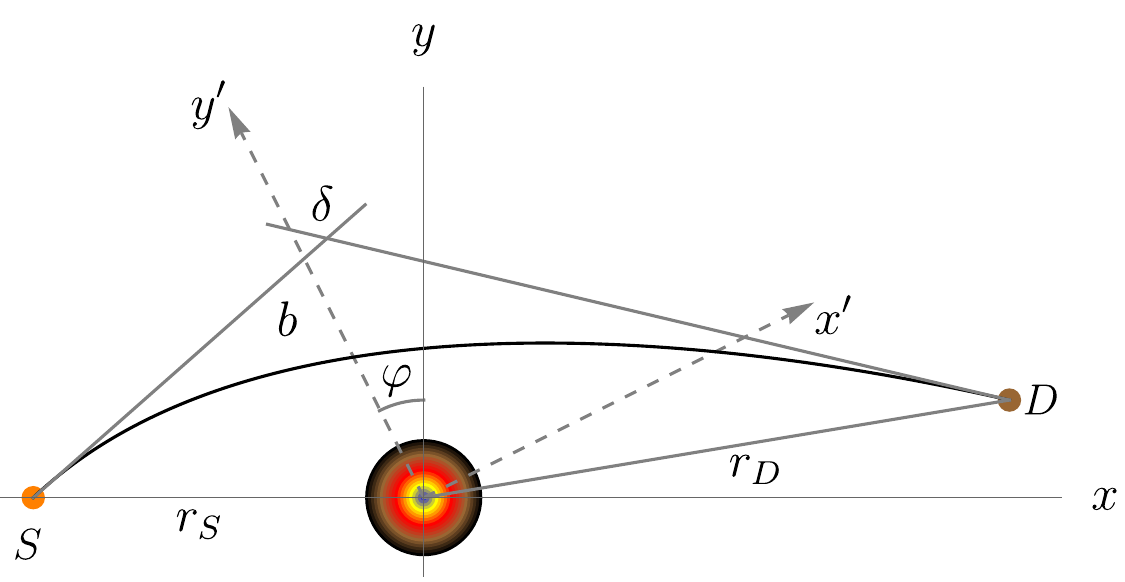}
    \caption{The diagram outlines how weak gravitational lensing modifies the paths followed by neutrinos as they travel through curved spacetime. The point labeled $D$ indicates the detector's position, and $S$ marks the region from which the neutrinos are emitted.}
    \label{laesnsssisnsg}
\end{figure}

When neutrinos propagate through a region influenced by a strong gravitational field produced by a massive source, their trajectories may deviate from radial paths due to gravitational lensing effects \cite{neu54}. As a result, multiple distinct routes can lead neutrinos to converge at the same detection point $D$, as depicted in Fig.~\ref{laesnsssisnsg}. This scenario requires a redefinition of the neutrino flavor state to include the superposition of all possible paths \cite{neu64,neu62,neu65,Shi:2024flw,neu56,neu63,AraujoFilho:2025rzh}:
\begin{align}
|\nu_{\alpha}(t_{D},x_{D})\rangle = N\sum_{i}\mathrm{U}_{\alpha i}^{\ast}
\sum_{p} e^{- \mathrm{i} \Phi_{i}^{p}}|\nu_{i}(t_{S}, x_{S})\rangle.
\end{align}

Here, each path taken by the neutrinos is identified by an index $p$. Since all these trajectories ultimately intersect at the detection point, the probability of detecting a flavor transition from $\nu_{\alpha}$ to $\nu_{\beta}$ results from the coherent combination of contributions from every possible route. This total transition probability is formulated as follows \cite{Shi:2024flw,neu56,neu62,neu63,neu64,neu65}:
\begin{align}
\label{nasndkas}
\mathcal{P}_{\alpha\beta}^{\mathrm{lens}} & = |\langle \nu_{\beta}|\nu_{\alpha}(t_{D}, x_{D})\rangle|^{2}\notag\\
& =|N|^{2}\sum_{i, j}\mathrm{U}_{\beta i}\mathrm{U}_{\beta j}^{\ast}\mathrm{U}_{\alpha j}\mathrm{U}_{\alpha j}^{\ast}\sum_{p, q}e^{\Delta\Phi_{ij}^{pq}}.
\end{align}
From this, the normalization constant takes the form:
\begin{align}
|N|^{2} = \left[\sum_{i}|\mathrm{U}_{\alpha i}|^{2}\sum_{p,q}\exp\left(-\mathrm{i}\Delta\Phi_{ij}^{pq}\right)\right]^{-1}.
\end{align}

The phase shift $\Delta\Phi_{ij}^{pq}$, previously introduced, is essential to determine the oscillation probability of neutrinos in the presence of gravitational lensing. This probability is shaped by various parameters, including the masses of the neutrino species, their mass--squared differences, and the characteristics of the curved spacetime generated by the black hole, as formulated in Eq.~(\ref{nasndkas}). The resulting structure is similar to that one observed in spherically symmetric spacetimes like the Schwarzschild metric \cite{neu53,Shi:2024flw,AraujoFilho:2025rzh} recently addressed in the literature.

The discussion now centers on how gravitational lensing affects the probability of neutrino flavor transitions, with a specific focus on the contribution from the Lorentz--violating parameter $\ell$. In such a lensing framework involving two--flavor neutrinos, the transition probability $\nu_{\alpha} \to \nu_{\beta}$ at the detector is computed by applying the weak-field approximation and considering the relative positioning of the source, the gravitational lens, and the detector \cite{neu53,neu54,neu55,Shi:2024flw,neu65}
\begin{align}
\label{asdPdadbd2}
\mathcal{P}_{\alpha\beta}^{\mathrm{lens}}
&=\left|N\right|^2\biggl\{2\sum_i\left|U_{\beta i}\right|^2\left|U_{\alpha i}\right|^2\left[1+\cos\left(\Delta b_{12}^2B_{ii}\right)\right]\notag\\
&\quad+\sum_{i\neq j}U_{\beta i}U_{\beta j}^*U_{\alpha j}U_{\alpha i}^*\notag\\
&\quad\times\left[\exp\left(-\mathrm{i}\Delta m_{ij}^2 A_{11}\right)+\exp\left(-\mathrm{i}\Delta m_{ij}^2 A_{22}\right)\right]\notag\\
&\quad+\sum_{i\neq j}U_{\beta i}U_{\beta j}^*U_{\alpha j}U_{\alpha i}^*\notag\\
&\quad\times\exp\left(-\mathrm{i}\Delta b_{12}^2B_{ij}\right)\exp\left(-\mathrm{i}\Delta m_{ij}^2A_{12}\right)\notag\\
&\quad+\sum_{i\neq j}U_{\beta i}U_{\beta j}^*U_{\alpha j}U_{\alpha i}^*\notag\\
&\quad\times\exp\left(\mathrm{i}\Delta b_{21}^2B_{ij}\right)\exp\left(-\mathrm{i}\Delta m_{ij}^2A_{21}\right)\biggr\}.
\end{align}

Eq.~(\ref{asdPdadbd2}) presents the transition probability as a combination of several terms, each enclosed in curly brackets and associated with particular configurations of mass indices and propagation paths. The term involving $i = j$ reflects contributions from a single mass eigenstate propagating independently, without interference. When the indices satisfy $i \neq j$ and $p = q$, the corresponding term represents interference effects resulting from different mass eigenstates acquired distinct phases while following the same trajectory. 

Additional contributions emerge when both the mass indices and the propagation paths differ ($i \neq j$, $p \neq q$). These terms are further separated based on the ordering of the path indices—whether $p < q$ or $p > q$—with each case evaluated individually due to asymmetries in the accumulated phase structure.

In scenarios where only two neutrino flavors are considered, the mixing framework simplifies considerably. The transformation between flavor and mass eigenstates is then described by a $2 \times 2$ unitary matrix, which depends entirely on a single mixing angle, denoted by $\alpha$ \cite{neu43}
\begin{align}
\label{U}
\mathrm{U}\equiv\left(\begin{matrix}
\cos\alpha&\sin\alpha\\
-\sin\alpha&\cos\alpha
\end{matrix}\right).
\end{align}

By inserting the mixing matrix defined in Eq.~(\ref{U}) into the general transition probability formula provided in Eq.~(\ref{asdPdadbd2}), one obtains the explicit expression for the oscillation probability corresponding to the flavor conversion process  
$\nu_e \rightarrow \nu_\mu$, as shown below
\begin{align}
\label{proballl}
\mathcal{P}_{\alpha\beta}^{\mathrm{lens}}
&=\left|N\right|^2\sin^{2}2\alpha\notag\\
&\quad\times\biggl[\sin^2\left(\dfrac{1}{2}\Delta m_{12}^2A_{11}\right)+\sin^2\left(\dfrac{1}{2}\Delta m_{12}^2A_{22}\right)\notag\\
&\quad+\dfrac{1}{2}\cos\left(\Delta b_{12}^2B_{11}\right)+\dfrac{1}{2}\cos\left(\Delta b_{12}^2B_{22}\right)\notag\\
&\quad-\cos\left(\Delta b_{12}^2B_{12}\right)\cos\left(\Delta m_{12}^2A_{12}\right)\biggr].
\end{align}

Taking into account the leptonic mixing matrix from Eq.~(\ref{U}) and the phase shifts associated with each distinct neutrino path, the expression for the normalization constant can be written as:
\begin{align}
\left|N\right|^2&=\biggl[2+2\cos^2\alpha\cos\left(\Delta b_{12}^2B_{11}\right)\notag\\
&\quad+2\sin^2\alpha\cos\left(\Delta b_{12}^2B_{22}\right)\biggr]^{-1}.
\end{align}


\section{Numerical investigation}

To investigate how neutrino oscillations manifest in the spacetime influenced by the black hole considered, it is necessary to evaluate the lensing probabilities given in Eq.~(\ref{proballl}). In the chosen $(x, y)$ reference frame, the lens occupies the origin, while the source and detector are positioned at distances $r_S$ and $r_D$ from this central point. To simplify the spatial configuration, a rotated frame $(x', y')$ is introduced by applying a transformation that rotates the original coordinates by an angle $\varphi$, yielding the relation \cite{neu53,Shi:2024flw}:  
\begin{align}
x' = x\cos\varphi + y\sin\varphi, \quad y' = -x\sin\varphi + y\cos\varphi .
\nonumber
\end{align}

Notice that by setting $\varphi = 0$ aligns the source, lens, and detector along the same linear axis within the plane, placing all three elements on a common straight--line configuration.

As established in Refs.~\cite{neu53,Shi:2024flw}, the gravitational bending of a neutrino’s trajectory is characterized by the deflection angle $\delta$, which is directly connected to the impact parameter $b$ through the following relation:
\begin{align}
\label{delta}
\delta \sim\frac{y_{D}'- b}{x_{D}'}=-P-\dfrac{4(1-\ell)^{\frac{3}{2}}M}{b},
\end{align}
with
\begin{align}
P\equiv\left(\sqrt{1-\ell}-1\right)\pi.
\end{align}

Eq.~(\ref{delta}) can be expanded to the first order with the parameter $\ell$:
\begin{align}
P+\dfrac{4(1-\ell)^{\frac{3}{2}}M}{b}\simeq\dfrac{4M}{b}-\dfrac{\pi\ell}{2}-\dfrac{6M\ell}{b}
\end{align}
In contrast to the deflection angle formula found in Ref.~\cite{yang2023static}, we have more precisely determined an extra coupling term $-\frac{6M\ell}{b}$ that holds significant physical importance: it reflects the nonlinear interaction between the gravitational mass and the Kalb--Ramond gravity by directly relating the Lorentz symmetry breaking parameter $\ell$ to the gravitational source mass $M$. 

Because the deflection angle $\delta$ depends on $\ell$ and $M$ only through the combination $M(1-\ell)^{3/2}/b$, varying the impact parameter $b$ changes merely an overall scale. Consequently, measurements of $\delta$ alone cannot disentangle $\ell$ from $M$. Breaking this degeneracy demands a second observable whose dependence on $\ell$ is not proportional to $1/b$—for instance, the Shapiro time delay or perihelion advance.

Placing the detector at coordinates $(x_{D}', y_{D}')$ within the rotated reference frame, and applying the relation $\sin\varphi = \frac{b}{r_S}$, the expression given in Eq.~(\ref{delta}) can be rewritten in the following form:
\begin{align}
\label{solve_b}
&\quad\left[by_D+bPx_D+4(1-\ell)^{\frac{3}{2}}Mx_D\right]\sqrt{1-\dfrac{b^2}{r_S^2}}\notag\\
&=b^2\left(\dfrac{x_D}{r_S}+1+\dfrac{yP}{r_S}\right)-4(1-\ell)^{\frac{3}{2}}M\dfrac{b_yD}{r_S}.
\end{align}

A comprehensive examination of neutrino oscillations in the presence of a Kalb--Ramond black hole is undertaken to assess the role played by the Lorentz--violating parameter $\ell$. This analysis contrasts the modified gravitational background with the standard Schwarzschild case, retrieved in the limit $\ell = 0$. The impact parameter is computed using Eq.~(\ref{solve_b}), which introduces an additional contribution labeled by $P$, arising naturally from the Kalb--Ramond correction and not present in the Schwarzschild scenario, as one should expect.

\begin{figure*}
\centering
\includegraphics[height=5cm]{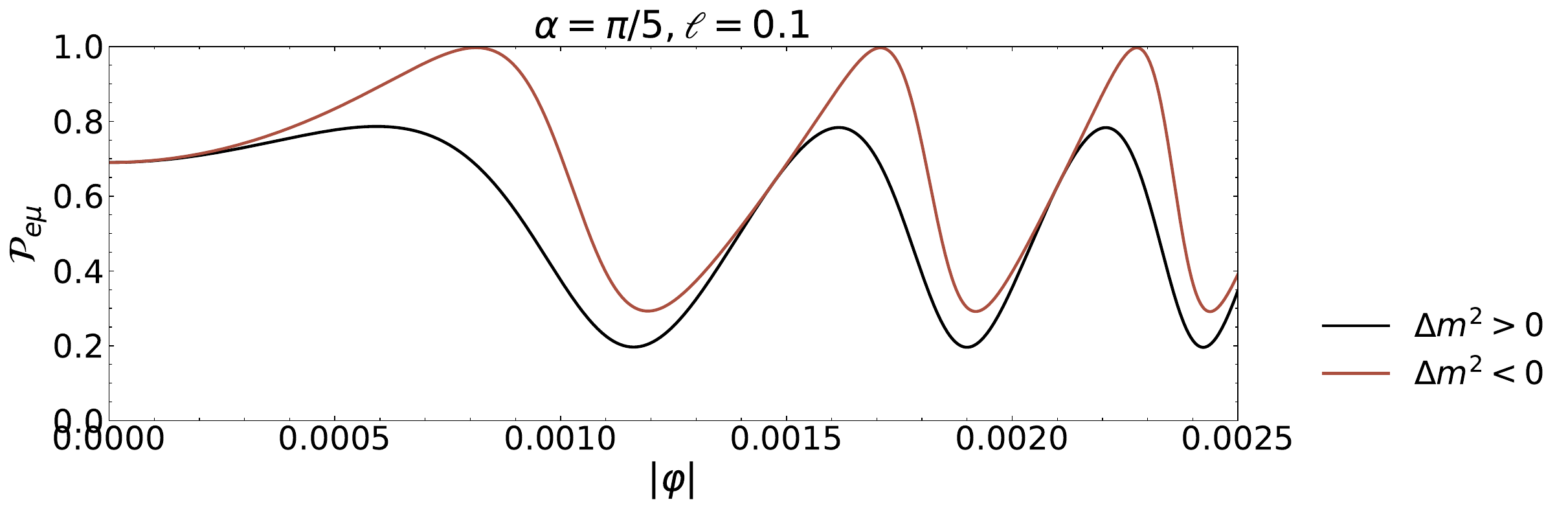}
\includegraphics[height=5cm]{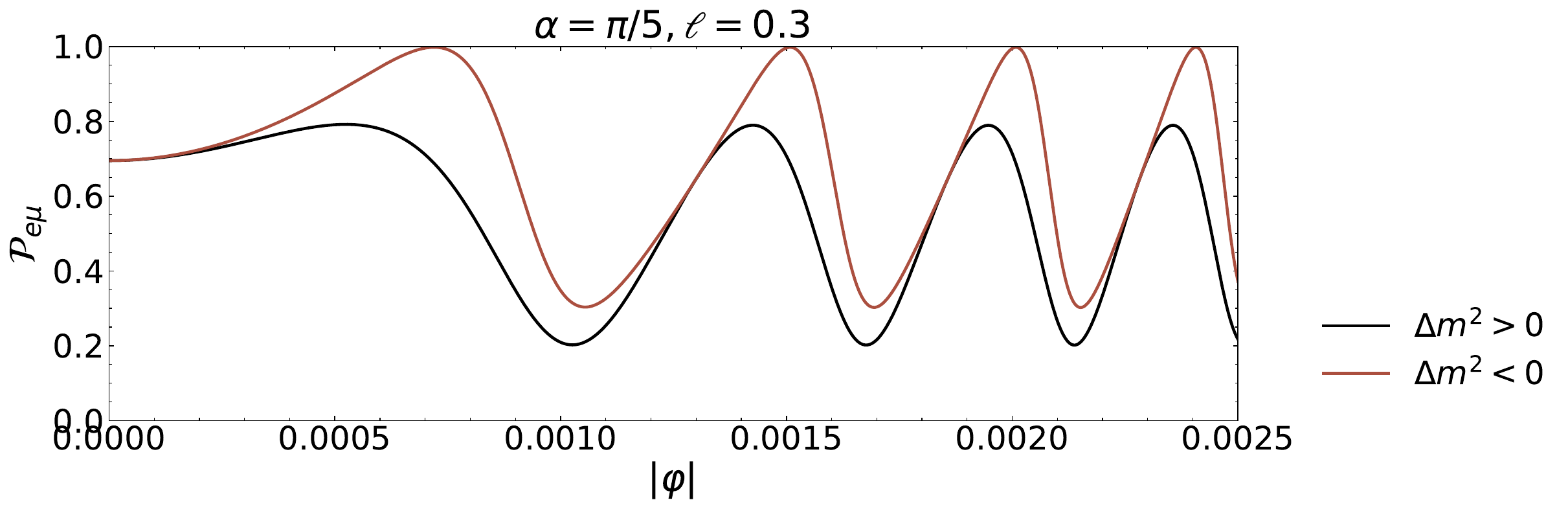}
\includegraphics[height=5cm]{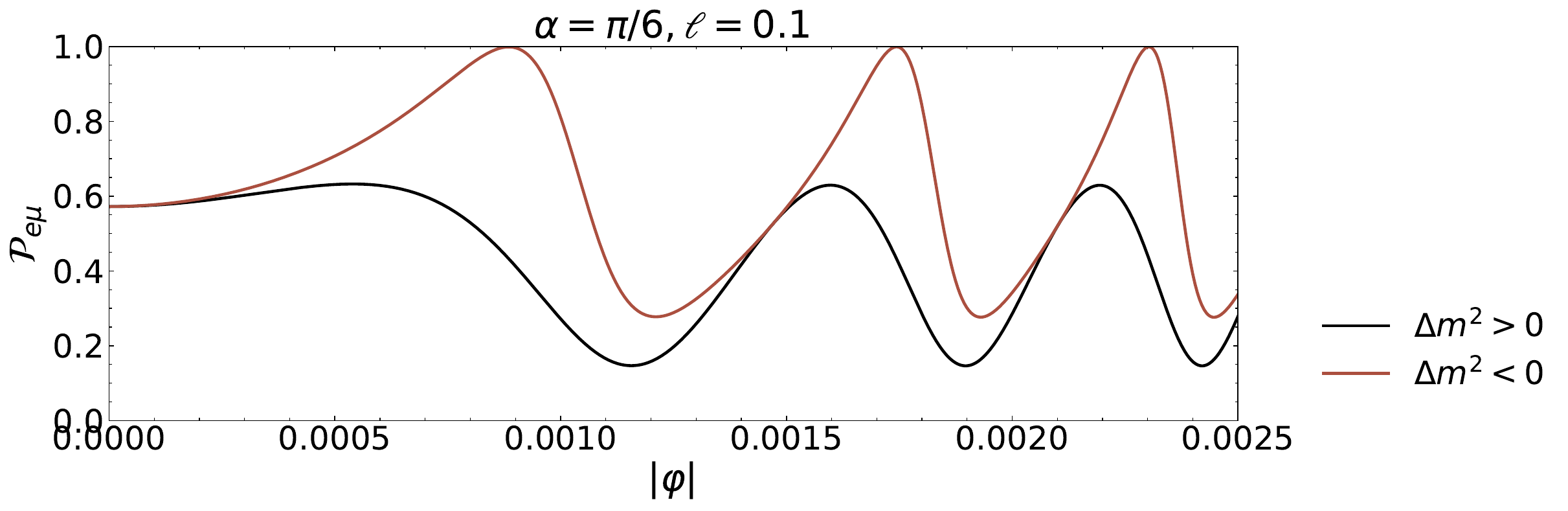}
\includegraphics[height=5cm]{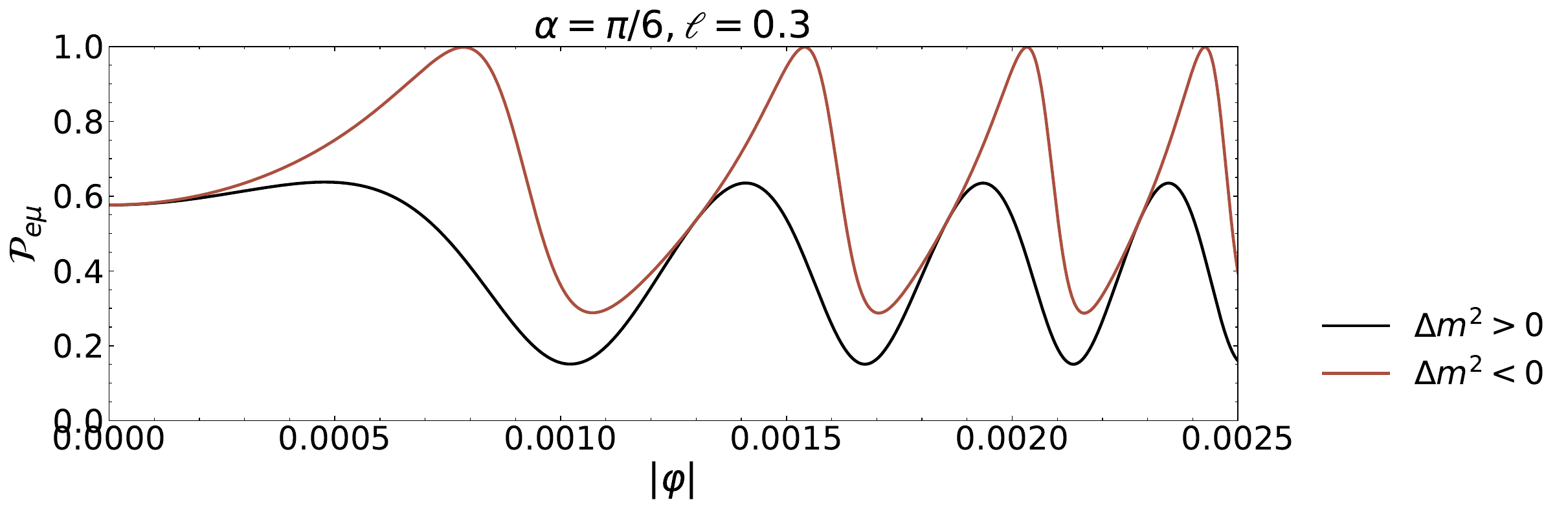}
\caption{\label{fig:prob1} The behavior of the conversion probability $\nu_e \to \nu_\mu$ is analyzed as a function of the azimuthal angle $\varphi$ for selected values of the Lorentz--violating parameter, specifically $\ell = 0.1$ and $0.3$. This investigation is carried out within a two--flavor oscillation scheme, considering both normal and inverted mass orderings, and employs mixing angles $\alpha = \pi/5$ and $\alpha = \pi/6$.}
\end{figure*}

\begin{figure*}
\centering
\includegraphics[height=5cm]{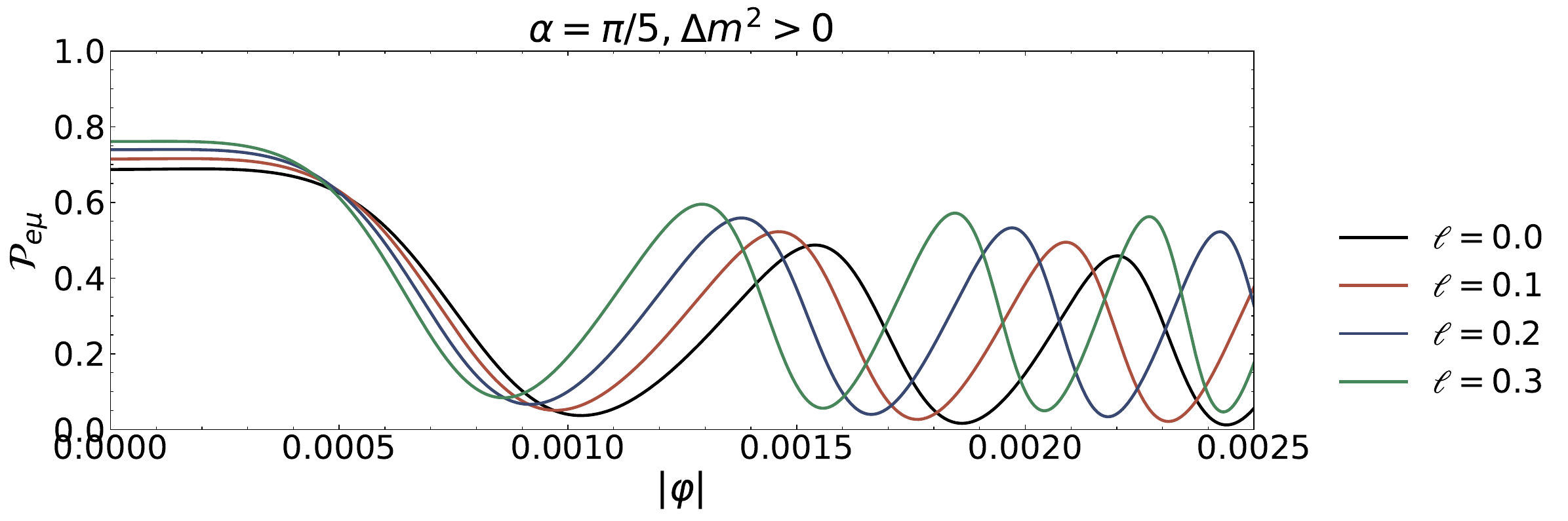}
\includegraphics[height=5cm]{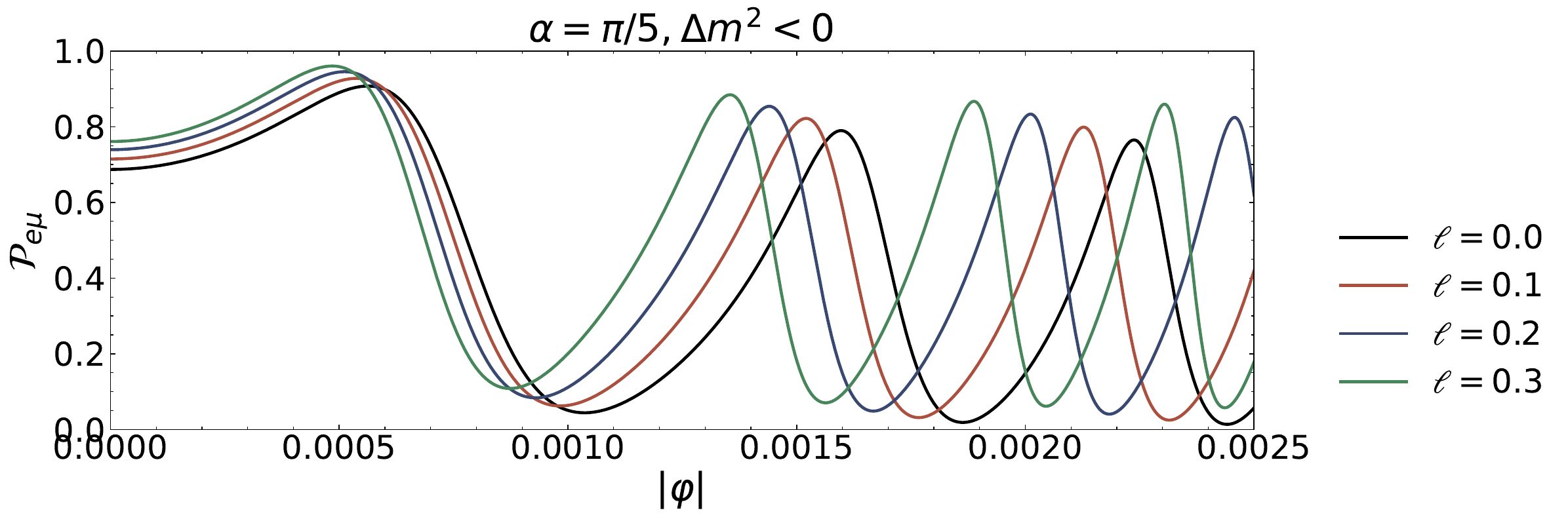}
\includegraphics[height=5cm]{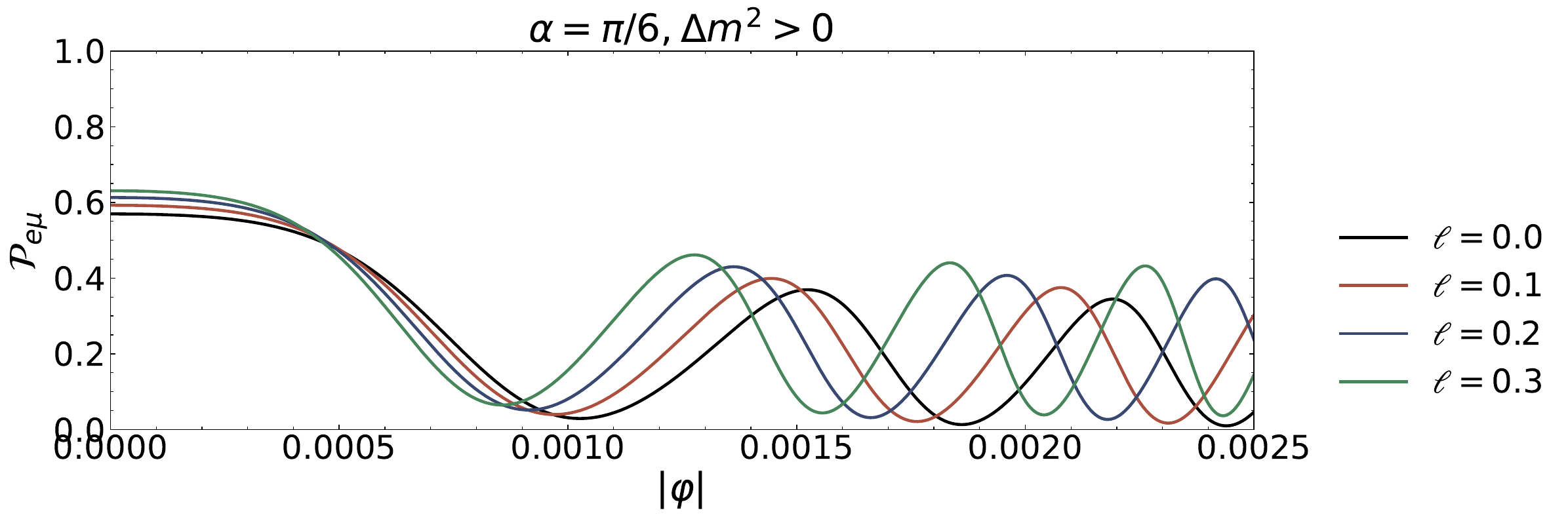}
\includegraphics[height=5cm]{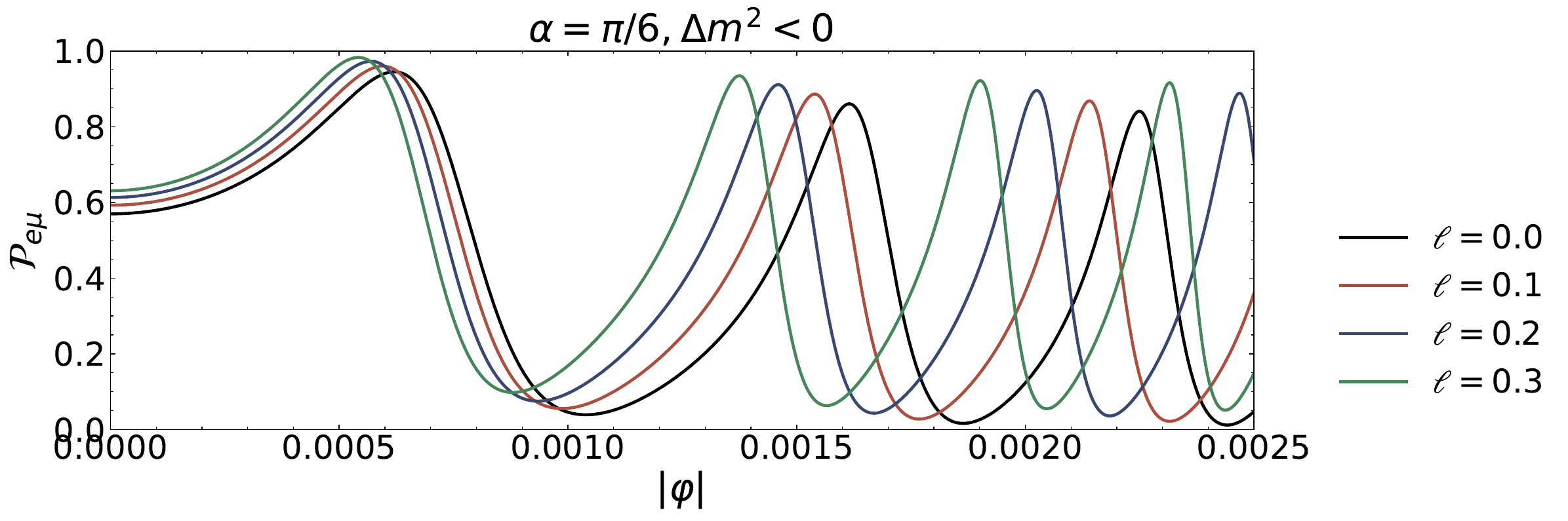}
\caption{\label{fig:prob2} The transition probability for $\nu_e \to \nu_\mu$ is evaluated as a function of the azimuthal angle $\varphi$, considering $\ell = 0.1$ and $0.3$. The analysis adopts a two--flavor oscillation model and accounts for both normal and inverted mass orderings, with mixing angles chosen as $\alpha = \pi/5$ and $\alpha = \pi/6$.}
\end{figure*}

\begin{figure*}
\centering
\includegraphics[height=5cm]{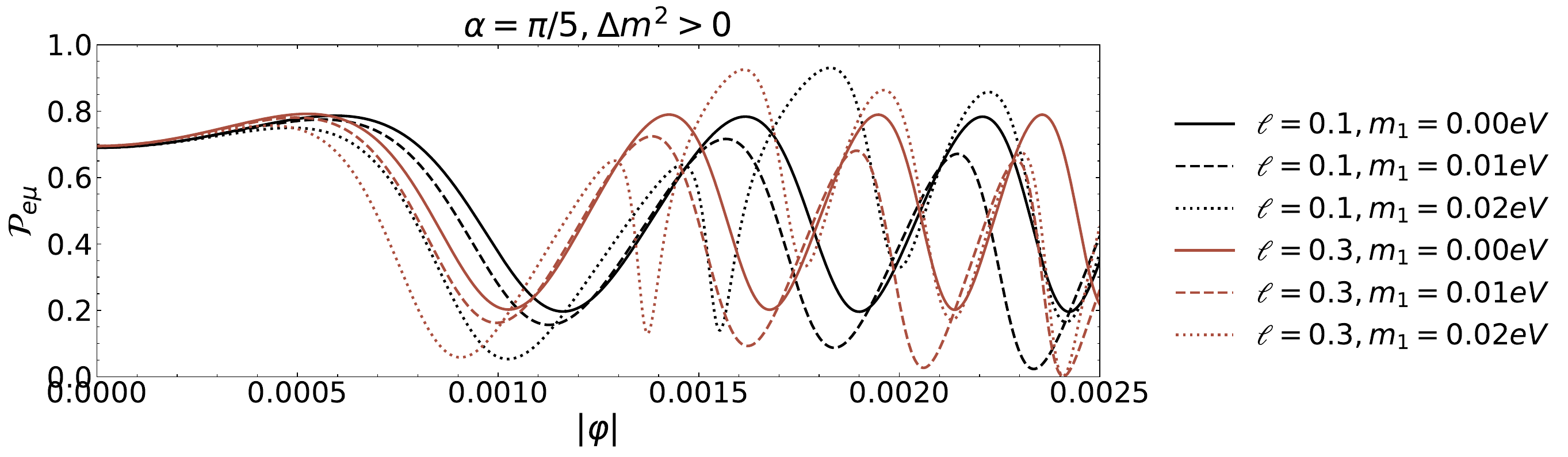}
\includegraphics[height=5cm]{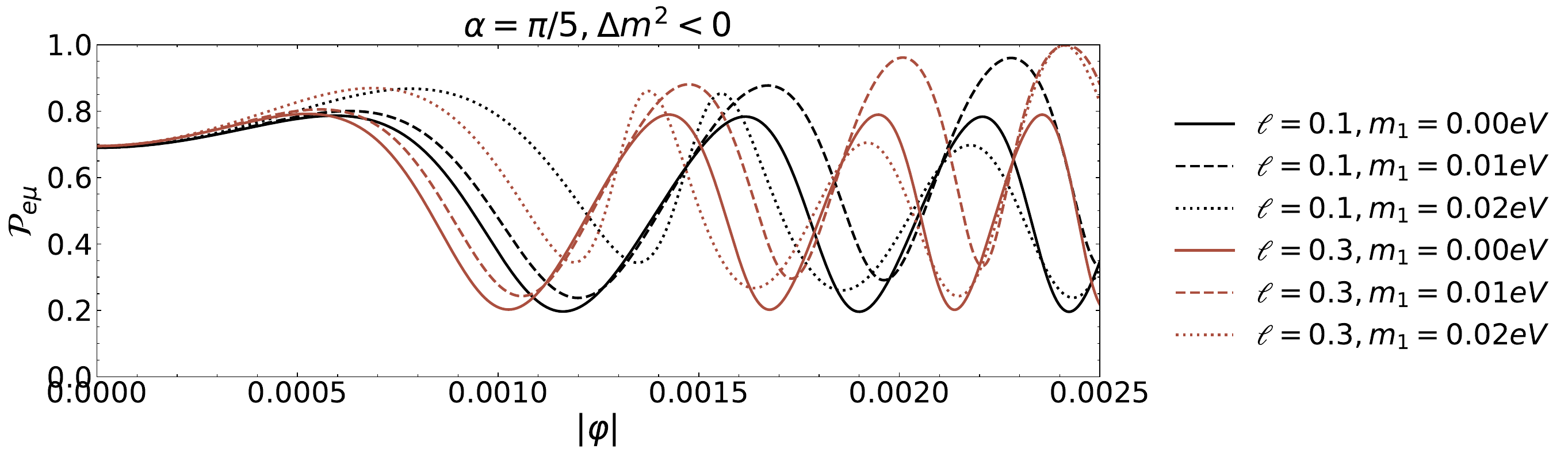}
\includegraphics[height=5cm]{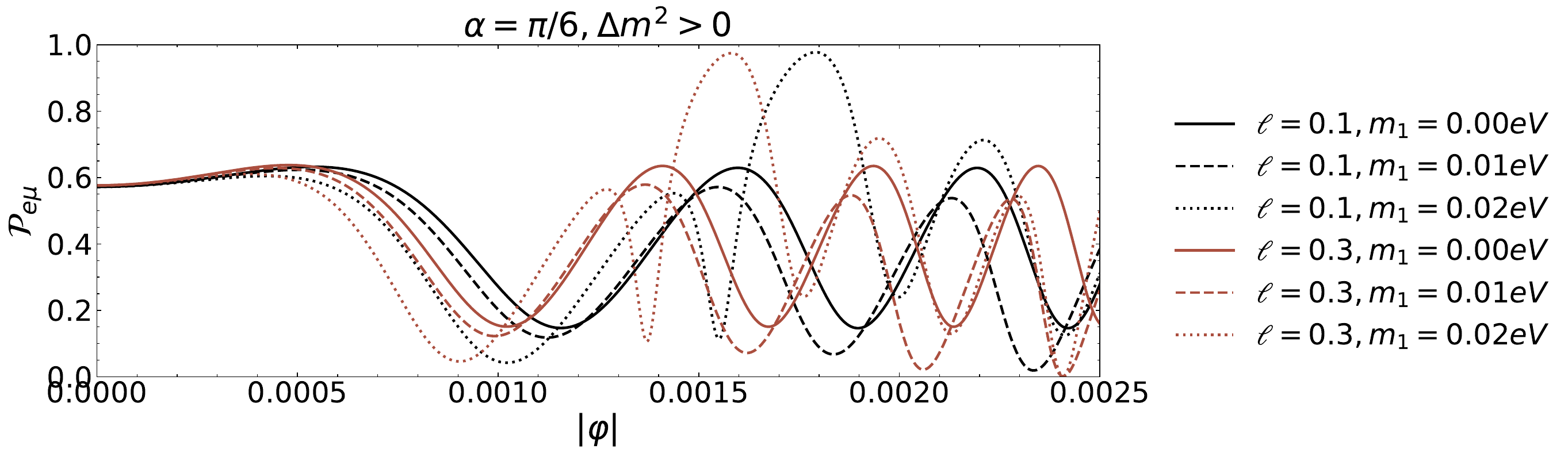}
\includegraphics[height=5cm]{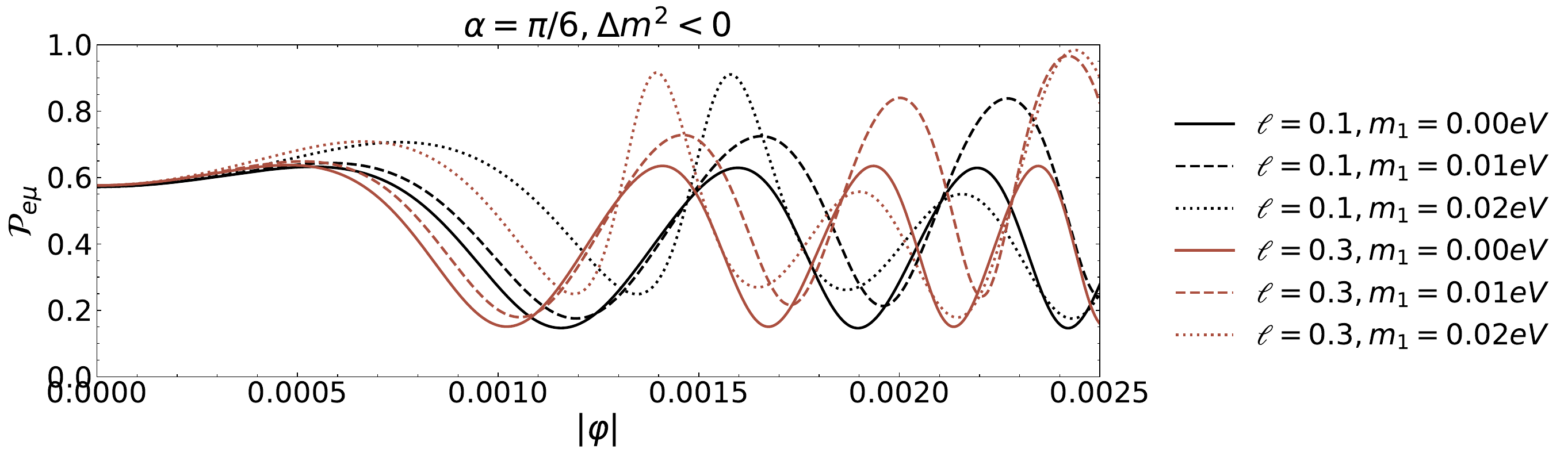}
\caption{\label{fig:prob3} The dependence of the neutrino oscillation probability on the azimuthal angle $\varphi$ is shown for both mass orderings: normal ($\Delta m^2 > 0$) and inverted ($\Delta m^2 < 0$). The curves correspond to two values of the Lorentz--violating parameter, with black representing $\ell = 0.1$ and red indicating $\ell = 0.3$. Different line styles denote distinct choices of the lightest neutrino mass: solid for $m_1 = 0\,\mathrm{eV}$, dashed for $m_1 = 0.01\,\mathrm{eV}$, and dotted for $m_1 = 0.02\,\mathrm{eV}$.}
\end{figure*}

Figs. ~\ref{fig:prob1}, \ref{fig:prob2}, and \ref{fig:prob3} display how the probability of neutrino flavor transitions varies with respect to the azimuthal angle $\varphi$. In the first two figures, the lightest neutrino mass is taken to be zero. Two mixing angles are considered: $\alpha = \pi/5$ and $\alpha = \pi/6$. The flavor conversion $\nu_e \rightarrow \nu_\mu$ is examined in Fig.~\ref{fig:prob1}, where black lines represent normal mass ordering ($\Delta m^2 > 0$), while red lines correspond to inverted ordering ($\Delta m^2 < 0$). The transition profiles for fixed values of $\ell$ reveal consistent structures across panels, and in every case, inverted ordering leads to higher transition probabilities compared to the normal configuration. These results demonstrate that both the sign and magnitude of $\Delta m^2$ significantly influence how gravitational lensing modifies the oscillation patterns.

To explore these effects in greater detail, Fig.~\ref{fig:prob2} compares transition probabilities by varying $\ell$, the mixing angle, and the sign of $\Delta m^2$. The observed trends reinforce the conclusions from Fig.~\ref{fig:prob1}, particularly regarding the impact of $\ell$. As $\ell$ increases, oscillations occur more rapidly, although the qualitative shape of the curves remains similar, in line with recent previous works \cite{swami2020signature,Shi:2024flw,neu53}.

The first half of Fig.~\ref{fig:prob2} focuses on the normal mass ordering under both values of $\alpha$, plotting results for $\ell = 0$ (black), $\ell = 0.1$ (red), $\ell = 0.2$ (blue), and $\ell = 0.3$ (green). The second half presents the inverted hierarchy for the same parameter sets. Across all configurations, an increase in $\varphi$ reduces the oscillation period and amplifies the probability amplitude, regardless of the mass ordering.

Fig.~\ref{fig:prob3} shifts focus to the influence of the lightest neutrino mass, examining cases where $m_1 = 0\,\mathrm{eV}$, $0.01\,\mathrm{eV}$, and $0.02\,\mathrm{eV}$, under both $\ell = 0.1$ and $\ell = 0.3$. These panels explore the normal hierarchy ($\Delta m^2 > 0$), while equivalent conclusions for the inverted case ($\Delta m^2 < 0$) are also established. The results show that the oscillation profile changes notably with the absolute mass scale, even when $\ell$ is fixed. The dependence on the azimuthal angle $\varphi$ persists across all panels, with outcomes influenced by both the sign of $\Delta m^2$ and the choice of mixing angle. Finally, comparing results for different $\ell$ values reveals pronounced differences in both amplitude and frequency.

Another relevant remark is that the smaller the values of $\ell$, the less pronounced the differences between them become, making it difficult to distinguish their effects. This has been recently observed in studies of neutrino behavior within bumblebee gravity, both in the standard metric formulation \cite{neutrinobumblebee} and in the metric--affine framework \cite{neutrinobumblebeemetricaffine}.

\section{Conclusion}

This study essentially explored how Lorentz symmetry breaking, introduced by a black hole solution in Kalb--Ramond gravity, affected neutrino behavior. The investigation focused on three central aspects: the energy release from neutrino–antineutrino annihilation, modifications to neutrino oscillation phases due to the background geometry, and alterations in flavor transition probabilities arising from gravitational lensing. In addition to the theoretical analysis, a numerical treatment was conducted to evaluate oscillation probabilities under both normal and inverted mass orderings, employing a two--flavor approximation for comparison.

In particular, the energy deposition rate from $\nu\bar\nu \to e^{+}e^{-}$ was significantly reduced: for a source with radius $R = 20\,$km and luminosity around $10^{53}\,$erg s$^{-1}$, the total energy output was cut by half for $\ell = 0.1$ and dropped by nearly an order of magnitude for $\ell = 0.3$, when compared to the Schwarzschild case. During propagation, each mass eigenstate acquired a phase given by $\Phi_k\simeq\frac{m_k^{2}}{2E_0}(r_D+r_S)\!\bigl[1-\frac{b^{2}}{2(1-\ell)r_Dr_S}+2(1-\ell)\frac{M}{r_D+r_S}\bigr]$, indicating that higher values of $\ell$ increased the oscillation length. Although our numerical analysis considered values in the range $\ell \sim 0.1$–$0.3$ to highlight the phenomenological patterns, future studies were expected to examine whether such effects remained detectable for values consistent with effective field theory frameworks, particularly in light of the stringent bounds on SME coefficients. Matching procedures, such as those that related $\ell$ to the isotropic limit of $s_{00}$ in the SME, might have helped clarify the physical viability of these scenarios.

Gravitational lensing also influenced flavor transition probabilities: larger values of $\ell$ led to shorter oscillation periods in the azimuthal angle and enhanced amplitudes. In all scenarios examined, inverted mass ordering yielded higher $\nu_e \to \nu_\mu$ conversion probabilities than the normal one. The absolute neutrino mass affected the oscillation pattern, and changed in $\ell$ produced noticeable variations in amplitude and frequency across different configurations.


\section*{Acknowledgments}
\hspace{0.5cm} A. A. Araújo Filho is supported by Conselho Nacional de Desenvolvimento Cient\'{\i}fico e Tecnol\'{o}gico (CNPq) and Fundação de Apoio à Pesquisa do Estado da Paraíba (FAPESQ), project No. 150891/2023-7.

\bibliographystyle{ieeetr}
\bibliography{main}

\end{document}